# Prototype matching: children's preference for forming scientific concepts

Wang Zhong[1] Zhang Yi[2] Jiang Yi[3]

[1](Beijing Doers Education Consulting Co.,Ltd, Beijing 101107,China)

[2](Beijing Fengtai District Fengtai First Primary School,Beijing 100071,China)

[3](Beijing Dongcheng District Dongsijiutiao Primary School, Beijing 100007, China)

**Abstract:** Inspired by a sample lesson, this paper studies and discusses children's preferences in learning scientific concepts. In a "Dissolution" lesson, one of the students took the demonstration experiment of "carmine dissolves in Water" demonstrated by the teacher as the prototype to judge whether a new phenomenon belongs to dissolution, instead of analyzing and judging the phenomenon by using the dissolution definition. Therefore, we propose a conjecture that "prototype matching" may be a more preferred way for children to learn concepts than thinking through inquiry experiment, analysis, deduction, etc. To this end, we conducted a targeted test on 160 fifth grade students (all of whom had learned this lesson) from a primary school in Beijing, and used goodness of fit test to statistically analyze the results. The results showed that: ① the Chi square of the general result is 73.865, P<0.001, indicating that children did have obvious prototype preference; ② We "tampered" some of the prototypes, that is, they looked like the prototypes that the teacher had told students, but they were actually wrong. However, the results showed that children still preferred these so-called "prototypes" (chi square is 21.823, P<0.001). Conclusion: ① Children have an obvious preference for "prototype matching" in scientific concept learning, which is not only obviously deviated from the current general understanding of science education that emphasizes discovery/inquiry construction, but also points out that there may be a priority relationship among various ways of concept organization (such as definition theory, prototype theory, schema theory, etc.). ② Children's preference for prototypes seems to be unthinking, and they will not identify the authenticity of prototypes, which is particularly noteworthy in front-line teaching.

## 1. Origin of research

This research originated from a science lesson of "dissolution". In this lesson, the teacher asked the students to guess what factors might accelerate the dissolution. One of the girls replied that putting ice in hot water is accelerated dissolution. However, from the definition of dissolution, the melting of ice in hot water obviously does not belong to dissolution, because the premise of dissolution must be that two different substances are mixed together (Baike.baidu, "DISSOLUTION"). So why does this student think that ice melts in hot water is dissolved? To this end, we conducted an after-class interview with this student (see "9. Appendix" of this article for details). We noticed the last video. In this video, we did not directly ask her why the phenomenon of ice immersion in hot water is dissolution, but imagined a phenomenon: mixing blue and yellow pigments together to become green. Does this phenomenon belong to dissolution? Through this question, we try to understand the student's understanding of the concept of "dissolution" in his brain, and then understand why this student has the above point of view. However, it was unexpected that this student's explanation of this hypothetical (she thought it was dissolution) was a little strange: instead of explaining and analyzing this phenomenon, she matched it with an experiment demonstrated by the teacher in that class, namely, the phenomenon that carmine dissolved in water. She thinks that this phenomenon is very similar to the demonstration experiment. Since the carmine is dissolution, naturally, this phenomenon (blue mix yellow become green) should also be dissolution.

Although this answer does not point to the original question, it leaves us a clue: it seems that we can see how this student understands the concept of "dissolution". If we understand this problem, the problem of ice and hot water will be solved. From this question and answer, we can clearly feel that the student's learning about the phenomenon of "dissolution" seems to follow "prototype matching", that is, if the carmine experiment is known in advance to be dissolution, then she will try to take the newly encountered phenomenon to match with this experiment. What can be matched is dissolution, and vice is not. In addition, the criteria for matching were set by the student himself in an instant.

If this conjecture is correct, it will bring considerable enlightenment to the existing science teaching, especially the concept teaching. At present, the domestic primary school scientific community is more advocating the concept construction theory of Vygotsky-Branda Landsdown, that is, a learning progressing (see "2. Views of previous studies on this issue" for details). However, the situation of the girl suggests that whether there are many modes of scientific concept learning? And is Vygotsky's model the favorite preference in children's learning?

## 2. Views of previous studies on this issue

At present, domestic science education mainly follows Vygotsky's concept construction theory. Lev Vygotsky, a former Soviet scholar, elaborated his concept in his book Thought and Language. He believes that, first of all, children will mix a series of phenomena through observation to form a unique compound: compound thinking. The second step is that children's brains should process these compounds to form a "chain compound" - the "purest type of compound thinking" (Lev Vigotsky, 2005, P141). Third, children will abstract and summarize these compounds, that is, form a pseudo concept. But this obscure abstraction is extremely unstable and vulnerable to the interference of concrete phenomena. The key to stabilizing it is language. Vygotsky believes that the addition of language to cognitive processes makes pseudo concepts eventually form into scientific concepts. Therefore, Vygotsky's concept of conceptual learning can be roughly summarized as a process: compound thinking → chain compound → pseudo concept → concept (Lev Vygotsky, 2005, P115-181).

Vygotsky's mode is very close with the process that students experience in inquiry experiments: because students also have to experience a process of peeling off the cocoon and finding the truth in inquiry experiments. Therefore, under the strong advocacy of some scholars (such as B. Landsdown of Harvard University), inquiry learning based on Vygotsky has become the mainstream view of learning in science education (Branda Landsdown, P.E. Blackwood, P.F. Bradwin, 2008). The latest Science Curriculum Standards for Compulsory Education issued by the Ministry of Education of PRC takes scientific inquiry as one of the most important learning methods, and it was written into the general goal of the course: "The core qualities of students to be cultivated in science courses... including scientific concepts, scientific thinking, inquiry practice, attitude and responsibility" "To form the awareness of scientific inquiry, understanding scientific inquiry is the main way to explore and understand nature, obtain scientific knowledge, and solve scientific problems" (Ministry of Education of PRC, 2022, P4-7).

However, some previous psychological studies have confirmed that human conceptual learning may follow a variety of ways, and prototype matching is an important one. For example, Rips et al. found that people responded more quickly to the sentence "robin is a bird" than to the sentence "turkey is a bird". In this regard, K M. Galotti believes that, for most people, robin is a typical bird, but turkey is not (Katheleen M. Galotti, 2017, P113). As Wang Shugen mentioned, "it (prototype matching) believes that the memory stored in the human brain is not a template corresponding to the external model, but a prototype. This model reflects the basic characteristics of a class of objects" (Wang Shugen, 2002). However, it is a pity that at present, the psychological circle does not seem to have much research on the priority of various ways of concept learning, nor does it point out which may be more preferred by children (previous preference research has focused more on the overall learning model, while the organization and preference of concepts and knowledges are relatively rare). Relevant research is more seen in animal experiments. For example, Thomas A. Daniel et al. conducted abstract concept learning training on young pigeons, and found that these pigeons have obvious preference effects (Thomas A. Daniel et al, 2016). However, this paper described pigeons' preference behavior for matching as "oddity", indicating that in the eyes of researchers, this concept learning method should not be a representative phenomenon.

In fact, some previous studies have shown that students have shown signs of seeking prototypes in science learning. For example, Ye Baosheng and Peng Xiang found in their research on the pre scientific concept of primary school students that primary school students have the characteristics of "outstanding and obvious characteristics" in concept learning. For example, when they studied the characteristics of children's textile fabrics, they found that 60% of students believed that the characteristics of textile fabrics were soft. In this regard, the article believes that: "Textile fabrics are things with comprehensive characteristics, but the proportion of students' understanding of the feature of 'softness' is far higher than other features... Pupils only have a clear understanding of their' softness', which indicates that they have a distinctive cognitive feature" (Ye Baosheng, Peng Xiang, 2018). Obviously, this "softness" has already been prototype. However, these signs have not attracted the attention of relevant domestic educational scholars, not only have they not been written into the previous curriculum standards, but also can hardly be found in the influential papers on "concept learning".

**3 Basic conjecture of this study**

Compared with Vygotsky's progressing construction method, "prototype matching" may be a more preferred way for children to learn scientific concepts.

**4. Experimental design**

4.1 Basic ideas

Because the study is about preference, this experiment adopts distributed fitting as the basic idea. That is, we take "students have no special preference in learning scientific concepts, but only make judgments based on calm analysis" as zero hypothesis $H_0$. And then let the participants (students) complete a series of choice questions (test tasks), all of which have only two options, A and B, one of which contains the prototype they have been exposed to; The other one is not——it requires careful thinking and analysis to determine whether it is correct. According to $H_0$, the test results should be random, so the probability of selecting the option with prototype should be close to 0.5 (when the total number of participants is enough). Finally, we use goodness of fit test to verify whether the actual observed frequency fits 0.5, and then judge whether $H_0$ is acceptable.

4.2 Research route

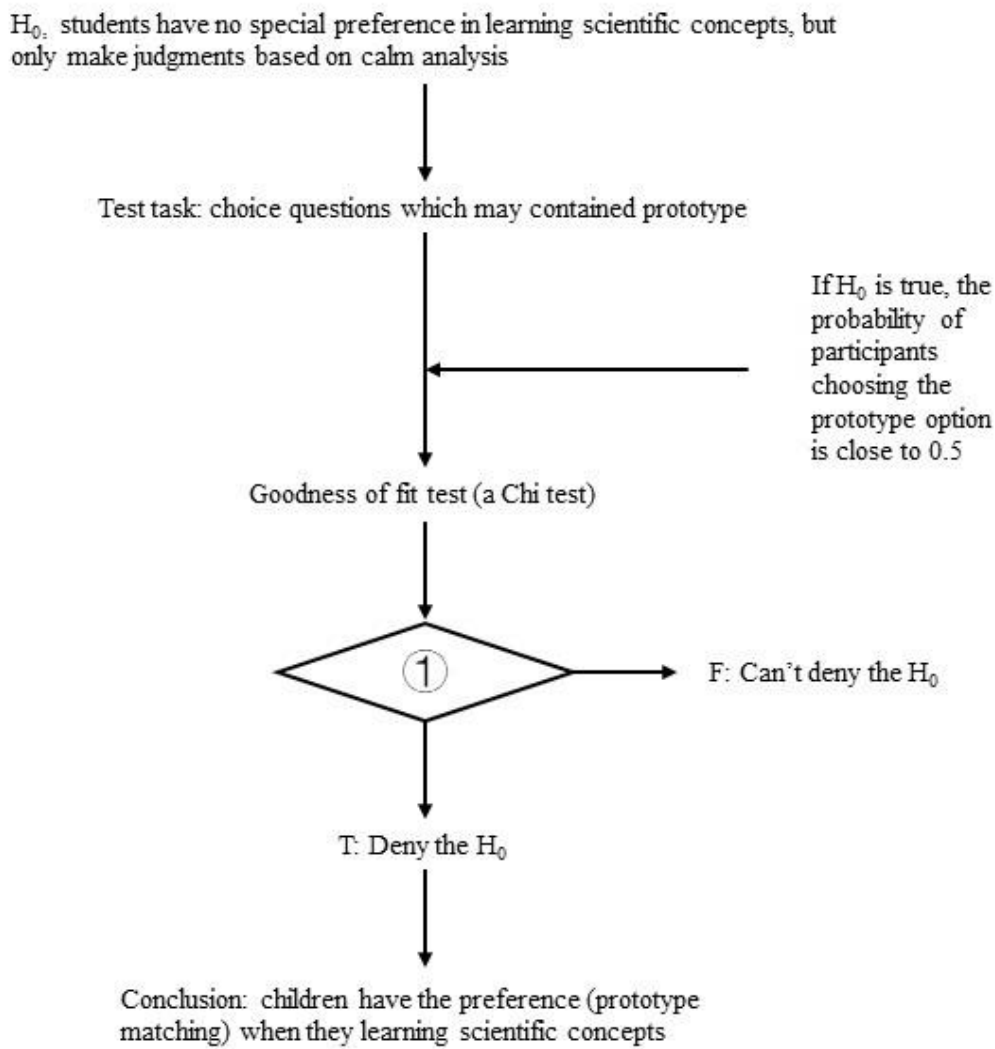

Figure 1

① Is there a significant difference between the actual observed frequency and 0.5 (P<0.05)?

4.3 Test task design

First of all, because the current knowledge content of biology, universe, earth science in primary school stage almost does not involve experiment and inquiry learning, but use the teaching method based on observation, we can't screen the students' preferences through this experiment, so the test task does not include the above content.

Secondly, in order to ensure that the prototype students of the test task have been exposed to it, the prototype options in all topics are based on the experiments (or pictures) in the textbook (Yu Bo, 2020). In order to highlight the experimental effect, we "tampered" some prototypes to make them look like the prototypes they have been exposed to, but after careful analysis, participants can find that they are actually wrong and cannot explain the knowledge points involved in this question.

Third, because there is often many typical experiments or models presented in textbooks, we cannot confirm which of them are the prototype of students, so we designed two sets of questionnaires (test tasks). The knowledge points involved in the two sets of questionnaires are identical, and the non-prototype options are also identical. The only difference is the options include prototypes.

See "9. Appendix" for questionnaires. Among them, the red font is the option with prototype, and the question number marked with "*" is our option of "tampered". The red font and "*" will not appear in the real questionnaire.

4.4 Design of statistical analysis method

We used goodness of fit test as the main statistical method of this study. The reason for choosing this method is that, first, as mentioned above, the research target of this experiment is preference, and second, it can "judge whether there is a significant difference between the expected frequency and the observed frequency" (Baike.baidu, "GOODNESS OF FIT TEST"). Therefore, it can be used to explain whether the deviation between the sample results of this experiment and 0.5 is caused by inevitable random errors.

Previous studies have also shown that goodness of fit test is very suitable for behavioral preference research. For example, Melanie et al. studied the preference of 1282 people in Germany for colorectal cancer screening through this method. Because screening is of great significance for early diagnosis and treatment of colorectal cancer, but it also has certain risks (Baike.baidu, "COLONOSCOPY"), so patients have a game based decision-making problem on whether to choose. Because of that, that paper examined the preference through goodness of fit test (Melanie Brinkmann MSc et al, 2022).

4.5 Effect size method design

As mentioned above, this study is based on chi square test (goodness of fit test), so we use Cramer's V correlation coefficient as the calculation method of the effect size. The calculation formula of this method is:

$$V = \sqrt{\frac{\chi^2}{n(k-1)}}$$

Where, n is the sample size, and k is the smaller of the number of rows and columns (Zheng Haomin, Wen Zhonglin, Wu Yan, 2011).

The size standard of the effect of this method is small effect V<0.1, medium weak effect V∈(0.1, 0.3], medium effect V∈(0.3, 0.5], large effect V>0.5). (Baike.baidu, EFFECT SIZE)

4.6 Sample size design

Since this study is based on goodness of fit test, which objectively requires a high sample size, we should especially avoid conclusion bias caused by too small sample size (the so-called "sparse data") (Hu Chunyan et al., 2021). For this purpose, we use a large sample size (n>30). At the same time, we require that the number of participants rejected should not exceed 5% (inclusive) of the total number of participants.

**5. Experiment preparation and implementation**

5.1 Participants

We selected 160 fifth-grade students from a primary school in Beijing as participants. The reason for choosing the fifth-grade is to ensure that they have learned all the knowledges in the questionnaire and have been exposed to the prototypes. All students have medium academic ability and good IQ. This experiment has passed the ethical review of local institutions, and guaranteed the participant's right to know to the greatest extent without exposing the purpose of the test.

5.2 Instruments

The paper questionnaires were used for data collection in this experiment, SPSS24 software was used for data statistics and analysis, and the analysis instrument was a Redmi Book Pro14 laptop.

5.3 Experimental process

This experiment includes two rounds in total. The first round is the test task described above. See 4.3, 5.1 and 5.2 of this paper for details of the participants and related preparations. In the distribution of test papers, the principle of random distribution is adopted, that is, two test papers are randomly mixed together by means of "shuffling" and randomly distributed to the participants by teachers who have not participated in this experiment.

The second round is an additional task, which aims to eliminate a potential competitive explanation: Are the options abandoned by most students those that students have not learned or will not (or even cannot understand)? Although the participants were in the fifth-grade, that is, the knowledge we chose should have been learned by students, because of the epidemic, many students had the experience of studying at home, and objectively it was difficult to guarantee the learning effect, so we added this task. We chose the questions which the participants no selected in the first round of test and redesigned the relatively detailed questions (except for a circuit question, all it was still choice questions) to see whether the students really can do it. In order to ensure that there is no consistency bias in the results (for example, doing the same questions again increases the experience and thinking time of the participants), we chose another class (38 students in total) with the same conditions as the participants. The students in this class have not experienced the first round of test tasks (that is, they haven't seen these questions), so their answers can reflect the average ability of the participants to deal with the first round of test tasks.

See "9. Appendix" for the second round of test questions (additional tasks).

## 6. Results and Analysis

6.1 Data preprocessing

Since most test tasks are options A and B, we must assign values to the answers before statistical analysis. The specific assignment method is as follows: if a subject chooses an option with prototype (no matter whether the option is A or B) when answering a question, we will assign the answer as 1; Otherwise, if the non prototype option is selected, the value is assigned as 0. The first question in the additional task is to use arrows to indicate the current direction. The correct answer should be 6 arrows. We calculate the number of arrows drawn by the subject and divide it by 6 to get the assignment of the question (if a subject draws 4 arrows correctly, we will assign $4 \div 6 \approx 0.67$). Since the positive and negative poles of the battery cannot be seen clearly, the current direction can be ignored, but all arrows drawn should follow the same direction. This assignment ensures that the data points to the point of "whether the probability of selecting the prototype is random", which is the focus of this study.

At the same time, we eliminated the answers of 5 subjects, mainly because of missed questions or multiple choices. The number of rejected products accounted for 3.13% of the total, which was in line with the sample size design (see 4.6 in this article for details).

The data after preprocessing are as follows:

| Volume 1 | | | | Volume 2 | | | |
|---|---|---|---|---|---|---|---|
| Q1 | Q2 | Q3 | Q4 | Q1 | Q2 | Q3 | Q4 |
| 1 | 1 | 1 | 1 | 1 | 1 | 0 | 1 |
| 1 | 0 | 1 | 1 | 1 | 1 | 0 | 1 |
| 1 | 0 | 0 | 0 | 0 | 1 | 1 | 1 |
| 0 | 1 | 1 | 1 | 1 | 0 | 1 | 0 |
| 1 | 1 | 1 | 1 | 1 | 1 | 1 | 1 |
| 1 | 1 | 0 | 1 | 1 | 1 | 1 | 1 |
| 1 | 1 | 1 | 1 | 1 | 1 | 0 | 0 |
| 1 | 1 | 0 | 0 | 1 | 0 | 0 | 1 |
| 0 | 1 | 0 | 0 | 1 | 1 | 0 | 1 |
| 0 | 0 | 1 | 0 | 1 | 1 | 0 | 1 |
| 1 | 0 | 1 | 1 | 0 | 1 | 1 | 0 |
| 0 | 1 | 1 | 1 | 1 | 0 | 1 | 0 |
| 0 | 1 | 1 | 1 | 0 | 1 | 1 | 0 |
| 0 | 0 | 1 | 1 | 1 | 1 | 1 | 1 |
| 0 | 1 | 0 | 0 | 1 | 1 | 0 | 0 |
| 1 | 1 | 0 | 1 | 1 | 1 | 1 | 1 |
| 0 | 1 | 0 | 0 | 0 | 1 | 0 | 0 |
| 1 | 1 | 0 | 0 | 0 | 1 | 1 | 1 |
| 1 | 1 | 1 | 1 | 1 | 1 | 1 | 1 |
| 1 | 1 | 1 | 1 | 1 | 1 | 1 | 1 |
| 0 | 0 | 1 | 0 | 1 | 1 | 0 | 0 |
| 0 | 0 | 1 | 0 | 0 | 1 | 1 | 0 |
| 0 | 0 | 1 | 1 | 1 | 0 | 0 | 0 |
| 0 | 1 | 1 | 1 | 1 | 1 | 1 | 1 |
| 1 | 1 | 1 | 1 | 1 | 1 | 1 | 1 |
| 1 | 1 | 1 | 1 | 1 | 1 | 1 | 1 |
| 0 | 0 | 1 | 1 | 1 | 0 | 1 | 1 |
| 0 | 0 | 1 | 1 | 1 | 1 | 1 | 0 |
| 0 | 0 | 1 | 0 | 1 | 1 | 1 | 0 |
| 1 | 0 | 1 | 0 | 1 | 0 | 1 | 1 |
| 0 | 0 | 1 | 0 | 1 | 1 | 1 | 0 |
| 1 | 0 | 0 | 1 | 0 | 0 | 1 | 1 |
| 0 | 0 | 0 | 1 | 1 | 0 | 1 | 1 |
| 0 | 1 | 0 | 1 | 0 | 1 | 1 | 1 |
| 1 | 1 | 1 | 1 | 1 | 1 | 0 | 0 |
| 1 | 1 | 1 | 1 | 0 | 1 | 1 | 1 |
| 1 | 1 | 1 | 1 | 1 | 0 | 1 | 1 |
| 1 | 1 | 0 | 1 | 1 | 0 | 0 | 1 |
| 1 | 1 | 1 | 1 | 1 | 1 | 1 | 0 |
| 1 | 1 | 1 | 1 | 1 | 0 | 1 | 1 |
| 1 | 1 | 1 | 1 | 1 | 1 | 0 | 1 |
| 1 | 0 | 1 | 1 | 1 | 1 | 1 | 0 |
| 1 | 1 | 0 | 0 | 1 | 0 | 0 | 1 |
| 1 | 0 | 0 | 0 | 0 | 0 | 1 | 0 |
| 1 | 0 | 1 | 0 | 1 | 1 | 1 | 0 |
| 1 | 1 | 1 | 1 | 0 | 0 | 1 | 1 |
| 1 | 1 | 0 | 1 | 0 | 1 | 1 | 1 |
| 1 | 1 | 0 | 1 | 0 | 1 | 1 | 0 |
| 1 | 1 | 1 | 1 | 1 | 1 | 1 | 0 |
| 1 | 1 | 1 | 1 | 1 | 1 | 1 | 0 |
| 1 | 1 | 1 | 1 | 1 | 0 | 1 | 0 |
| 0 | 1 | 1 | 0 | 1 | 1 | 1 | 1 |
| 1 | 1 | 1 | 0 | 0 | 1 | 1 | 1 |
| 1 | 1 | 1 | 1 | 0 | 1 | 1 | 0 |
| 1 | 1 | 1 | 0 | 1 | 1 | 0 | 1 |
| 1 | 1 | 1 | 1 | 1 | 1 | 1 | 1 |
| 1 | 0 | 0 | 0 | 1 | 1 | 1 | 1 |
| 1 | 0 | 0 | 1 | 1 | 0 | 1 | 1 |
| 1 | 0 | 0 | 0 | 0 | 1 | 1 | 0 |
| 1 | 1 | 0 | 1 | 1 | 1 | 1 | 0 |
| 1 | 0 | 0 | 0 | 1 | 1 | 0 | 1 |
| 1 | 0 | 1 | 0 | 1 | 0 | 1 | 0 |
| 1 | 1 | 1 | 0 | 1 | 0 | 1 | 0 |
| 1 | 0 | 1 | 1 | 1 | 1 | 1 | 1 |
| 1 | 0 | 0 | 0 | 1 | 1 | 1 | 0 |
| 1 | 1 | 1 | 0 | 1 | 1 | 1 | 0 |
| 1 | 1 | 1 | 0 | 1 | 1 | 1 | 1 |
| 1 | 0 | 1 | 1 | 0 | 1 | 1 | 1 |
| 0 | 0 | 1 | 0 | 1 | 1 | 0 | 0 |
| 1 | 0 | 0 | 0 | 1 | 1 | 1 | 1 |
| 0 | 1 | 0 | 1 | 0 | 1 | 1 | 1 |
| 1 | 0 | 1 | 1 | 1 | 1 | 1 | 1 |
| 0 | 0 | 0 | 0 | 1 | 1 | 1 | 1 |
| 1 | 1 | 1 | 0 | 0 | 1 | 0 | 1 |
| 1 | 1 | 0 | 0 | 1 | 1 | 0 | 1 |
| 1 | 0 | 0 | 1 | 0 | 1 | 1 | 0 |
| —— | —— | —— | —— | 1 | 1 | 1 | 1 |
| —— | —— | —— | —— | 1 | 1 | 0 | 1 |
| —— | —— | —— | —— | 0 | 1 | 0 | 0 |

Table 1: Test Task Results

| Q1 | Q2 | Q3 | Q4 |
|---|---|---|---|
| 0.50 | 1 | 1 | 1 |
| 1.00 | 1 | 1 | 0 |
| 1.00 | 1 | 1 | 0 |
| 1.00 | 1 | 1 | 0 |
| 1.00 | 1 | 1 | 0 |
| 1.00 | 1 | 1 | 0 |
| 1.00 | 1 | 1 | 0 |
| 0.83 | 1 | 1 | 1 |
| 0.50 | 1 | 0 | 1 |
| 1.00 | 0 | 1 | 0 |
| 1.00 | 1 | 1 | 0 |
| 1.00 | 1 | 0 | 1 |
| 1.00 | 1 | 1 | 1 |
| 1.00 | 1 | 1 | 1 |
| 1.00 | 1 | 1 | 1 |
| 1.00 | 1 | 0 | 1 |
| 1.00 | 1 | 1 | 0 |
| 1.00 | 1 | 1 | 1 |
| 1.00 | 1 | 1 | 1 |
| 1.00 | 1 | 1 | 1 |
| 1.00 | 1 | 1 | 1 |
| 1.00 | 1 | 0 | 1 |
| 1.00 | 1 | 1 | 1 |
| 1.00 | 1 | 1 | 1 |
| 1.00 | 1 | 1 | 1 |
| 1.00 | 1 | 1 | 1 |
| 1.00 | 1 | 1 | 0 |
| 1.00 | 1 | 0 | 1 |
| 1.00 | 1 | 1 | 1 |
| 1.00 | 1 | 1 | 1 |
| 1.00 | 1 | 1 | 1 |
| 0.50 | 1 | 0 | 1 |
| 1.00 | 1 | 1 | 1 |
| 1.00 | 1 | 1 | 1 |
| 1.00 | 1 | 1 | 1 |
| 1.00 | 1 | 1 | 1 |
| 1.00 | 1 | 1 | 1 |
| 1.00 | 1 | 1 | 1 |

Table 2: Results of additional tasks

6.2 Excluding Competitive Interpretation

As described in 5.3 of this article, in order to prevent competitive interpretation - that is, the reason why these options are not selected is likely to be because we cannot do/understand the topic - we added a round of tests. The main content of this round of test is to select those options that are less selected by the subjects in the test task separately, and make a separate proposition to see whether the students with average learning ability of the subjects can answer these questions.

Table 3 shows the accuracy of 38 students' additional tasks (rounded to two decimal places). It can be seen that the average correct rate of students with the same academic level is 0.88, that is, as long as they think carefully, they will completely make these choices. Therefore, the competitive interpretation is basically excluded.

| Q1 | Q2 | Q3 | Q4 | Average |
|---|---|---|---|---|
| 0.96 | 0.97 | 0.84 | 0.74 | 0.88 |

Table 3

6.3 Inspection results and analysis

6.3.1 Overall statistical results

Table 4-6 shows the results after goodness of fit test, of which Table 4 is the descriptive statistics. It can be seen that the chi square value of the overall test task is 73.865, with a significance of $P<0.001$, and there is no cell with the expected frequency lower than 5. The correlation coefficient of Cramer's V, $V \approx 0.35$, has a medium effect. It is easy to see that the subjects have obvious prototype preference in their choice, and this preference has very obvious statistical significance.

| Cases | Average | SD | Min | Max |
|---|---|---|---|---|
| 620 | 0.6726 | 0.46965 | 0.00 | 1.00 |

Table 4

|  | Measured cases | Expected cases | SR |
|---|---|---|---|
| 0 | 203 | 310 | -107 |
| 1 | 417 | 310 | 107 |
| Sum | 620 |  |  |

Table 5

| Test statistic | |
| --- | --- |
| $\chi^2$ | 73.865 * |
| df | 1 |
| Progressive significance | 0.000 |

\* The expected frequency of 0 cells (0.0%) is less than 5. The minimum expected cell frequency is 310.0.

Table 6

6.3.2 the questions which "be tampered"

Although it can be seen from the data that the participants chose many prototype options, is it because those options are also correct? For this reason, we specially select the statistical results of those "tamper" topics with the "prototype" option (that is, the so-called "prototype" is actually wrong and cannot explain the scientific phenomenon referred to in the topic). Table 7-9 shows the statistical results of those topics. The effect size of Cramer's V≈0.31, has a medium effect. Similar to the approximate results shown in Table 4-6, the subjects still had obvious prototype preference. More importantly, although there are pitfalls in these questions, it seems that the subjects did not pay attention to this.

| Cases | Average | SD | Min | Max |
| --- | --- | --- | --- | --- |
| 231 | 0.6537 | 0.47683 | 0.00 | 1.00 |

Table 7

| | Measured cases | Expected cases | SR |
| --- | --- | --- | --- |
| 0 | 80 | 115.5 | -35 |
| 1 | 151 | 115.5 | 35 |
| Sum | 231 | | |

Table 8

|  | Test statistic |
| --- | --- |
| $\chi^2$ | 21.823 * |
| df | 1 |
| Progressive significance | 0.000 |

* The expected frequency of 0 cells (0.0%) is less than 5. The minimum expected cell frequency is 115.5.

Table 9

6.3.3 Behavioral difference of "having been involved"

What's more, do the subjects have obvious behavioral differences in answering the two types of questions, "Have you ever done anything?"? If they exist, it is likely that these problems affect the students (of course, it also means that the students are aware of these tricks), otherwise, they do not. In this regard, we used independent sample t-test to examine the differences of the subjects in answering the two questions.

|  | Cases | **Average** | SD | σ |
| --- | --- | --- | --- | --- |
| "be tampered" | 389 | 0.6838 | 0.46559 | 0.02361 |
| Never "be tampered" | 231 | 0.6537 | 0.47683 | 0.03137 |

Table 10

| | Levene test | | Independent t-test | | | | | 95% CI | |
| --- | --- | --- | --- | --- | --- | --- | --- | --- | --- |
| | F | Significance | t | df | **Significance (two tailed)** | Average interpolation | SE interpolation | Lower | Lower |
| EV* | 2.271 | 0.132 | 0.772 | 618 | 0.440 | 0.03012 | 0.03902 | -0.04651 | 0.10676 |
| UnEV | | | 0.767 | 474.079 | 0.443 | 0.03012 | 0.03926 | -0.04702 | 0.10727 |

*"EV" means the equal variance. Abbreviations are used because of insufficient table space*

Table 11

It can be seen from the above table that there is no significant difference in the data, which indicates that the subjects have no evidence to prove that they are aware of these tricks when answering the question of whether they have "cheated". From the data, it seems that the subjects are likely to have similar behaviors on these two types of questions.

To sum up the data analysis, we believe that the zero hypothesis $H_0$ can basically be denied, that is, from the data point of view, students have an obvious preference for prototype matching when identifying and judging concepts. More persuasively, even though the so-called "archetypes" were wrong, the subjects still stubbornly chose them. This seems to reflect that the subjects' choice of prototype is instantaneous and intuitive, because it is not difficult to find these traps as long as they think a little.

**7. Discussion**

7.1 Another evidence from "clinical"

The above results remind us of a case evidence several years ago. In 2018, we studied and analyzed a classroom teaching case. In this case, the teacher should explain the knowledge about the factors that affect the swing speed, so first let the students know the classical pendulum model. However, based on the idea of "coming from life", the teacher did not directly show the appearance of the pendulum, but started to guide students from the pendulum in life (such as the pendulum, swing, etc.), trying to let students construct the concept of "pendulum" through these. However, one of the students did not finally establish the concept of pendulum. He always believed that the pendulum should look like a swing (the pendulum explained in this lesson is a typical simple pendulum in physics). We interviewed the student after class: without prompting, the student was asked to draw the experiment in the class just now. We can clearly see (see the figure below for details) that the pendulums painted by this student have obvious characteristics of swings, which indicates that the pendulums constructed by this student are like swings (Wang Zhong, Miao Yingying, 2021).

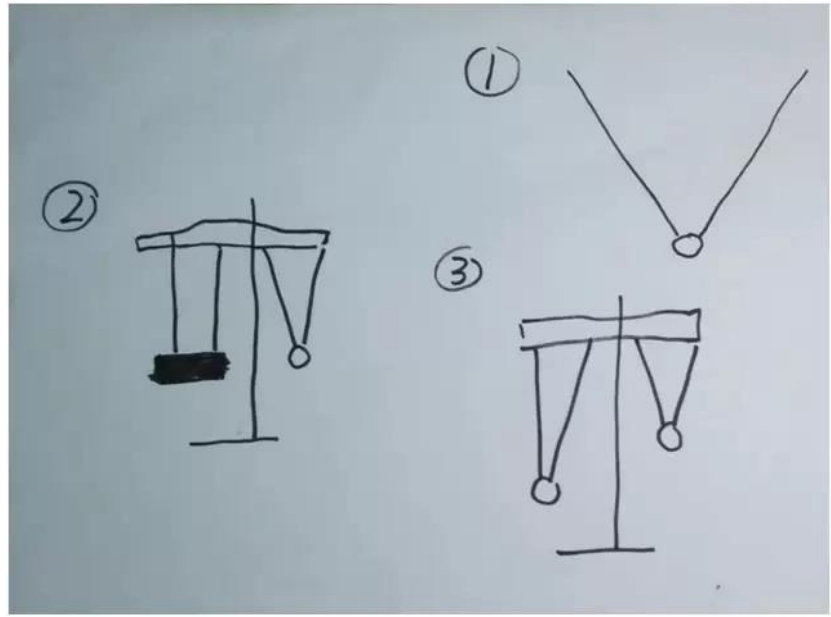

Figure 2

How to explain this case? If you carefully observe this picture, it is not difficult to find that compared with the simple pendulum (a cycloid) in physics, the so-called "pendulum" in this picture obviously has the characteristics of swing (especially ②). Therefore, from the perspective of this article, we can draw a logical explanation: when the teacher talked about the swing, the student unconsciously took the swing as the prototype, and then matched it with the shape of the pendulum. The result was that the match was not good, so he firmly denied that the pendulum was a pendulum. It can be said that this case provides an indirect evidence for this study from the front-line classroom.

7.2 Priority of concept formation

The greater impact of this article may lie in revealing the priority of conceptual organization.

For the formation of concepts, cognitive psychology generally includes "Classical view of concepts", "Prototype view of concepts", "Exemplar view of concepts", "Schema/script view of concepts" and "Knowledge based view of concepts" (Catherine Gallotty, 2017, P117-124). These viewpoints infer how the concept is formed from different angles, and put forward corresponding evidence. However, previous studies have revealed that any of these statements can not explain all the problems of concept formation, and even there will be a phenomenon of "jagged" understanding of concepts between different views. For example, Roach et al. found after investigation that most Americans believe that robins and sparrows are more like birds than chickens, penguins and ostriches, but these birds should have equal status in the classical view. Obviously, the classical view cannot explain the phenomenon that some birds are "more like birds" than others, and it seems more appropriate to explain this phenomenon with the prototype view and the example view (Catherine Galotti, 2017, P117). Therefore, it is obvious that different viewpoints (if they are all correct) are related to each other, and even have priorities.

Unfortunately, this does not seem to have received the attention it deserves. Most of the studies on concept preference in the literature focus on which concept the subjects prefer under the same conditions, rather than different learning (formation) methods of the same concept. For example, Olli J. Loukola et al. found in their research on the competition between great tits and banded flycatchers in nesting site selection that banded flycatchers are very sensitive to the size of the symbol on the nest of great tits (the symbol pre added by researchers on the nest of great tits), and even show a deliberate replication of this feature (during nesting), while they are not sensitive to factors such as shape (Olli J. Loukola et al. 2022). As another example, Thomas A. Daniel's research on young pigeons above also focuses on their preference for different color JPEG images (Thomas A. Daniel et al, 2016).

But obviously, this problem is also a very valuable one. The author has seen more than once signs that students are trying to find prototypes: for example, when looking for circuit problems (especially the circuit diagram on the test paper), they will deliberately look for what is not like a "circle" (the appearance of a simple circuit in the textbook), rather than measuring it through the current angle; Or the students who first learn the expansion and contraction of gas will explain the phenomenon that the gas expands and breaks the soap bubble when heated as "hot air will rise" (this phenomenon is very similar to the hot air balloon experiment they have seen in the classroom). It should be noted that these two examples (and the example from which this study originated) can be identified by reasoning and judgment, but students still use prototypes (even incorrect prototypes). This may mean that children will "package" things and principles requiring complex processing into prototypes in a certain way for processing, thus reducing the cognitive resources consumed by processing. Therefore, the study of concept preference is undoubtedly of great significance to the explanation of children's concept learning.

Furthermore, this problem may have another value for pure cognition: it is likely to help explain people's instantaneous decision-making problems. Instantaneous decision-making exists in many life situations, especially in purchase behavior. Previous research on this issue was basically based on the perspective of the "hypothesis of rational man", that is, the hypothetical person still analyzed relevant factors to establish decisions even in an instant. For example, Ji Xu and others built the production and operation model of the chemical industry based on the cost-benefit evaluation of INTERNET (Ji Xu, Zhu Lijia, 2003). For another example, Yuan Qian established an instantaneous uncertainty decision-making model based on the estimation of risk threshold and decision making on the purchase of venture capital (Yuan Qian, 2007). However, this paper reveals that in instantaneous decision-making, people are likely to be irrational, and their decision-making behavior is likely to be based on some "shortcuts" (such as trying to match the prototype).

However, obviously, due to insufficient attention and less relevant evidence, this hypothesis still needs more support from follow-up studies.

7.3 Deficiencies of this document

The biggest shortcoming of the methodology of this paper is that the core issues of this paper are not explored from the perspective of reaction time.

Many previous researches on prototype matching are based on reaction time, such as the robin and turkey examples listed in 2. For another example, Carol Conrad found that subjects did not react more slowly to the sentence "A shark can move" than to the sentence "A fish can move" or "Animal can move". This result is contrary to the Hierarchical semantic network model of semantic memory, because according to this theory, "shark" is a subclass of "fish", and it should be more laborious to call it naturally than "fish" or even "animal" (Catherine Garlotti, 2017, P111-112). It is easy to understand that if the activation of a concept has a preference, then activation according to preference should be faster than through complex analysis and reasoning. Therefore, reaction time should be a more ideal indicator.

Many previous researches on prototype matching are based on reaction time, such as the robin and turkey examples listed in 2. For another example, Carol Conrad found that subjects did not react more slowly to the sentence "A shark can move" than to the sentence "A fish can move" or "Animal can move". This result is contrary to the Hierarchical semantic network model of semantic memory, because according to this theory, "shark" is a subclass of "fish", and it should be more laborious to call it naturally than "fish" or even "animal" (Catherine Garlotti, 2017, P111-112). It is easy to understand that if the activation of a concept has a preference, then activation according to preference should be faster than through complex analysis and reasoning. Therefore, reaction time should be a more ideal indicator.

However, affected by the epidemic, we were unable to enter schools all the time, and naturally we were unable to bring reaction time equipment into schools and carry out relevant experiments. Therefore, the conclusion of this paper only serves as indirect evidence of this problem.

## 8. Full text conclusion

First, children have a clear preference for "prototype matching" in scientific concept learning, which is not only a clear deviation from the current general understanding of science education, which emphasizes discovery/inquiry construction, but also points out that various ways of concept organization (such as definition theory, prototype theory, schema theory, etc.) may have a priority relationship. Second, children's preference for archetypes seems to be unthinking, and they will not identify the authenticity of archetypes, which is particularly noteworthy in first-line teaching.

## 9. Appendix

All experimental data, test questions (including additional tasks) and initial post class interview videos of this paper have been uploaded to the Scientific Data Bank (ScienceDB). Please click the link to download or view them:

https://www.scidb.cn/en/detail?dataSetId=e3bddcd1754b422ba5870fef94346c13